\documentstyle[preprint,eqsecnum,aps]{revtex}
\begin{document}
\begin{center}
{\bf HIROTA BILINEAR FORMALISM AND SUPERSYMMETRY}
\end{center}
\begin{center}
A. S. Carstea{\footnote {E-mail: acarst@theor1.theory.nipne.ro}}
\end{center}
\begin{center}
{\it Institute of 
Physics and Nuclear 
Engineering, Dept. Theor. Physics, Magurele, Bucharest}
\end{center}
\vskip 1cm

\begin{abstract}
Extending the gauge-invariance principle for $\tau$ functions of 
the standard bilinear formalism to the supersymmetric case,
we define ${\cal N}=1$ supersymmetric Hirota operators.  
Using them, we bilinearize SUSY nonlinear evolution equations. 
The super-soliton solutions and extension to SUSY sine-Gordon are 
also discussed. As a quite strange paradox 
it is shown that the Lax integrable SUSY KdV of Manin-Radul-Mathieu
equation does not possess 
N super-soliton solution for $
N \geq 3$ for arbitrary parameters. Only for a particular
choice of them the N super-soliton solution exists.  
\end{abstract}

\section{Introduction}
Supersymmetric integrable systems constitute a very interesting subject
and as a consequence a number of well known integrable equations 
have been generalized into supersymmetric (SUSY) context.
We mention the SUSY 
versions of sine-Gordon \cite{di vecchia}, \cite{chaichain}, 
KP-hierarchy \cite{manin}, 
KdV \cite{manin},\cite{mathieu}, Boussinesq \cite{yung} etc.
We also point out that there are 
many generalizations related to the number ${\cal N}$
of fermionic independent variables. 
In this paper we are dealing with the
${\cal N}=1$ superspace.

So far, many of the tools used in standard 
theory have been extended to this
framework, such as 
B\"acklund transformations \cite{chaichain}, prolongation
theory, hamiltonian formalism \cite{oevel}, 
grasmmannian description 
\cite{ueno}, $\tau$ functions \cite{medina}, Darboux
transformations \cite{m}. The physical interest in 
 the study of these systems
have been launched by the seminal 
paper of Alvarez-Gaume et. al \cite{alvarez}
about the partition function and 
super-Virasoro constraints of 2D quantum supergravity.
Although the $\tau$ function theory in the 
context of SUSY pseudodifferential
operators was given for the SUSY KP-hierarchy \cite{ueno}, 
the bilinear formalism 
for SUSY equations was very little investigated.
We mention here the algebraic approach using the
representation theory of affine Lie super-algebras in the papers of 
Kac and van der Leur \cite{kac}, Kac and Medina\cite{kacmed}
the super-conformal field theoretic approach of LeClair \cite{leclair}. 
Anyway in these articles the bilinear hierarchies are not related
to the SUSY hierarchies of nonlinear equations.

This paper which is an extended version of \cite{fane} we consider a 
direct approach to SUSY equations in a ${\cal N}=1$ superspace 
rather than hierarchies 
namely extending the
gauge-invariance principle of $\tau$ functions for classical Hirota
operators. Our result generalize the Grammaticos-Ramani-Hietarinta
\cite{grammaticos} theorem, to SUSY case and we find 
${\cal N}=1$ SUSY Hirota bilinear operators. With these operators one can
obtain SUSY-bilinear forms for SUSY KdV equation of Mathieu \cite{mathieu}
and also it allows bilinear 
forms for certain SUSY extensions of Sawada-Kotera-Ramani \cite{sw}, 
Hirota-Satsuma \cite{satsuma}, KdV-B \cite{beck},
Burgers, mKdV and ${\cal N}=2$ -superspace SUSY 
sine-Gordon equations \cite{tsy}. 
Also the gauge-invariance principle allows to study the SUSY 
multisoliton solutions as exponentials of linears.
A very interesting fact 
which happens is that SUSY KdV equation of Mathieu does not have 
3 supersoliton solution for arbitrary choice of solitary waves although
it possesses Lax pair \cite{mathieu}. 
Only for special combination of 
parameters the equation admits N soliton solutions.
Although this seems 
to be a quite strange paradox, 
Liu and Manas \cite {m}, found also 
super-soliton solutions for SUSY KdV equation in terms of 
pfaffians only for certain wave parameters. 
This fact shows that Hirota integrability and 
Lax integrability are different in the SUSY context.

The paper is organized as follows. In section II the standard
bilinear formalism is briefly discussed. 
In section III supersymmetric versions
for nonlinear evolution equations are presented and in section IV 
we introduce the super-bilinear formalism.
In the last section we shall present the bilinear form for SUSY KdV-type
equations, super-soliton solutions and several comments about extension
to ${\cal N}=2$ SUSY sine-Gordon equation.

\section{Standard bilinear formalism}

The Hirota bilinear operators were introduced as an antisymmetric extension
of the usual derivative \cite{hir}, because of their usefulness for the 
computation of multisoliton solution of nonlinear evolution equations. 
The bilinear operator 
${\bf D}_{x}=\partial_{x_{1}}-\partial_{x_{2}},$
acts on a pair of functions (the so-called "dot product") antisymmetrically:
\begin{equation}
{\bf D}_{x}f\bullet g
=(\partial_{x_{1}}-\partial_{x_{2}})
f(x_{1})f(x_{2})|_{x_{1}=x_{2}=x}=f'g-fg'.
\end{equation}
The Hirota bilinear formalism has been instrumental in the derivation
of the multisoliton solutions of (integrable) nonlinear evolution equations.
The first step in the application is a dependent variable transformation
which converts the nonlinear equation into a quadratic form. This
quadratic form turns out to have the same 
structure as the dispersion relation
of the linearized nonlinear equation, 
although there is no deep reason for that.
This is best understood if we 
consider an example. Starting from paradigmatic
KdV equation
\begin{equation}
u_{t}+6uu_{x}+u_{xxx}=0,
\end{equation}
we introduce the substitution $u=2\partial_{x}^2\log F$ and obtain after
one integration:
\begin{equation}
F_{xt}F-F_{x}F_{t}+F_{xxxx}F-4F_{xxx}F_{x}+3F_{xx}^2=0,
\end{equation}
which can be written in the following condensed form:
\begin{equation}
({\bf D}_{x}{\bf D}_{t}+{\bf D}_{x}^4)F\bullet F=0.
\end{equation}
The power of the bilinear formalism lies in the fact that for multisoliton 
solution $F$'s are polynomials of exponentials. Moreover it displays also
the interaction (phase-shifts) 
between solitons. In the case
of KdV equation the multisoliton solution has the following form:
\begin{equation}
\label{N}
F=\sum_{\mu=0,1}\exp{(\sum_{i=1}^{N}\mu_{i}
\eta_{i}+\sum_{i<j}A_{ij}\mu_{i}\mu_{j})},
\end{equation}
where $\eta_{i}=k_{i}x-k_{i}^3 t+\eta_{i}^{(0)}$ and $exp{A_{ij}}=
(\frac{k_{i}-k_{j}}{k_{i}+k_{j}})^2 $ which is the phase-shift from
the interaction of the soliton $"i"$ with the soliton $"j"$.

This picture can be generalized to any bilinear equation of the form
\begin{equation}\label{gen}
P({\bf D}_{\vec x})F \bullet F=0 
\end{equation}
where $P$ is any polynomial and ${\vec x}=(t,x,y,...)$.
The 1 soliton solution is 
$F=1+e^{\eta}$, where $\eta={\vec k}{\vec x}+const.$ 
This solution holds if
$$P({\vec k})=0.$$ 
This is a condition on the 
parameters ${\vec k}$ of $\eta$ and is called the 
{\it dispersion relation}. 
If the parameter space is n-dimensional then the above equation defines 
an $n-1$ dimensional submanifold called {\it dispersion manifold}.

Hirota ansatz for 2 soliton solution is 
$$F=1+e^{\eta_1}+e^{\eta_2}+A_{12}e^{\eta_1+\eta_2}$$
where $\eta$'s are defined 
before and $A_{12}$ are function of ${\vec k}_1$ and ${\vec k}_2$
each one giving the 
coordinates of some point in the dispersion manifold. Substituting 
the ansatz in the 
equation (\ref{gen}) and taking into 
account the dispersion relation we find
\begin{equation}\label{it}
A_{12}P({\vec k}_1+{\vec k}_2)+P({\vec k}_1-{\vec k}_2)=0.
\end{equation}
Generally speaking, 
the great majority of 
bilinear equations possess 2 soliton solution but only
{\it integrable} equations 
possess 3 or more soliton solutions. The form of the N soliton
solution is,
$$F= \sum_{\mu=0,1}\exp{(\sum_{i=1}^{N}\mu_{i}
\eta_{i}+\sum_{i<j}A_{ij}\mu_{i}\mu_{j})}$$
More about the bilinear equations are in \cite{hieta}

A very important observation (which motivated the present paper)
is the relation of the physical field $u=2\partial_{x}^2\log{F}$ of KdV
equation with the Hirota function $F$: the gauge-transformation 
$F\rightarrow e^{px+\omega t}F$ leaves $u$ invariant. This is a general 
property of all bilinear equations. Moreover, one can define the Hirota
operators using the requirement of gauge-invariance. 
Let's introduce a general
bilinear expression, 
\begin{equation}
A_{N}(f,g)=\sum_{i=0}^{N}c_{i}(\partial_{x}^{N-i} f)(\partial_{x}^{i} g)
\end{equation}
and ask to be invariant under the gauge-trasformation:
\begin{equation}
A_{N}(e^{\theta}f,e^{\theta}g)=e^{2\theta}A_{N}(f,g)
\quad \theta=kx+\omega t+...(linears).
\end{equation}
Then we have the following,\cite{grammaticos}

{\it Theorem: $A_{N}(f,g)$ is gauge-invariant if and only if} 
$A_{N}(f,g)={\bf D}_{x}^{N}f\bullet g$ i.e. 
$$c_{i}=c_{0}(-1)^{i}
\left(  \begin{array}{c}
        N\\i
        \end{array}     \right)$$
and $c_{0}$ is a constant and the brakets represent binomial coefficient.

We must point out that in the whole paper the natural number N 
(which will denote number of solitons, number of terms, exponents etc.) 
is different
from ${\cal N}$ which is related to supersymmetry or superspace.
\section{Supersymmetry}

The supersymmetric extension of a nonlinear 
evolution equation (KdV for instance) refers to a system
of coupled equations for a bosonic $u(t,x)$ and a 
fermionic field $\xi(t,x)$ which reduces to the
initial equation in the limit 
where the fermionic field is zero (bosonic limit).
In the classical context, 
a fermionic field is described by an anticommuting
function with values in an {\it infinitely} generated Grassmann algebra.
However, supersymmetry is not just a coupling of a bosonic field to a 
fermionic field. It implies a transformation (supersymmetry invariance)
relating these two fields
which leaves the system invariant.
In order to have a mathematical formulation of these concepts we have to 
extend the classical space $(x,t)$ to a larger space (superspace) 
$(t,x,\theta)$  
where $\theta$ is a Grassmann variable and 
also to extend the pair of fields $(u,\xi)$ to a larger
fermionic or bosonic superfield $\Phi(t,x,\theta).$ 
In order to have nontrivial extension for KdV we choose $\Phi$ to 
be fermionic, having the expansion
\begin{equation}
\Phi(t,x,\theta)=\xi(t,x)+\theta u(t,x).
\end{equation}
The ${\cal N}=1$ SUSY means that we have
only one Grassmann variable $\theta$ and 
we consider only space supersymmetry invariance
namely $x\rightarrow x-\lambda\theta$ 
and $\theta\rightarrow \theta+\lambda$
($\lambda$ is an anticommuting parameter). 
This transformation is generated
by the operator 
$Q=\partial_{\theta}-\theta\partial_{x},$ 
which anticommutes
with the covariant derivative 
$D=\partial_{\theta}+
\theta\partial_{x}$ (Notice also that $D^2=\partial_{x}$). 
Expressions written in terms of the covariant derivative and the superfield
$\Phi$ are manifestly supersymmetric invariant.
%\newpage
Using the superspace formalism one can 
construct different supersymmetric extension of 
nonlinear equations. Thus the integrable (in the sense of Lax pair) 
variant of ${\cal N}=1$ SUSY KdV is \cite{manin}\cite{mathieu}
\begin{equation}\label{skdv}
\Phi_{t}+D^6 \Phi+3D^2(\Phi D\Phi)=0,
\end{equation}
which on the components has the form
\begin{eqnarray}
u_{t}&=&-u_{xxx}-6uu_{x}+3\xi\xi_{xx}\nonumber\\
\xi_{t}&=&-\xi_{xxx}-3\xi_{x}u-3\xi u_{x}.
\end{eqnarray}
Another integrable variant of 
SUSY KdV equation which is very important in applications to
supersymmetric matrix models is SUSY KdV-B equation \cite{beck}, namely
\begin{equation}
\Phi_{t}+D^6 \Phi+6D^2\Phi D\Phi=0,
\end{equation}
leads to a somewhat trivial 
system in which the fermionic fields decouple from the 
bosonic equation which reduces then to the usual KdV.

We shall discuss also the following supersymmetric equations, although
we do not know if it is completely integrable in the sense
of Lax pair ($\Phi$ is also a fermionic superfield).

\begin{itemize}
\item ${\cal N}=1$ SUSY Sawada-Kotera-Ramani,
\begin{equation}\label{sskr}
\Phi_{t}+D^{10} \Phi
+D^2(10 D\Phi D^4 \Phi+5 D^5 \Phi\Phi+15(D\Phi)^2 \Phi)=0.
\end{equation}

\item ${\cal N}=1$ SUSY Hirota-Satsuma (shallow water wave)
\begin{equation}
D^4 \Phi_{t}+\Phi_{t} D^3\Phi+2D^2 \Phi D\Phi_{t}-D^2 \Phi-\Phi_{t}=0
\end{equation}

\item ${\cal N}=1$ SUSY Burgers
\begin{equation}
\Phi_t+\Phi D\Phi_{x}+\Phi_{xx}=0
\end{equation}

\end{itemize}

A very important equation from the physical consideration is the
SUSY sine-Gordon. We are going to consider the version studied by
Kulish and Tsyplyaev \cite{tsy}. There are other integrable versions
of SUSY sine-Gordon emerged from algebraic procedures \cite{inami}. 
In this case one needs two Grassmann variables 
$\theta_{\alpha}$ with $\alpha=1,2$
and the supersymmetry transformation is
$$x^{'\mu}=x^{\mu}-i\bar\lambda\gamma^{\mu}\theta, \quad 
\theta_{\alpha}^{'}=\theta_{\alpha}+\lambda_{\alpha}, \mu=1,2.$$
Here, $\lambda_{\alpha}$ is the anticommuting spinor 
parameter of the transformation and 
$\bar\lambda=(\lambda^1, \lambda^2)$,
$\lambda^{\alpha}=\lambda_{\beta}(i\sigma_{2})^{\beta\alpha}$, 
$\gamma^{0}=i\sigma_{2}$, 
$\gamma^{1}=\sigma_{1}$, 
$\gamma^{5}=\gamma^{0}\gamma^{1}=\sigma_{3}$.
We use the metric $g^{\mu\nu}=diag(-1,1)$ and $\sigma_{i}$ are the Pauli 
matrices.
The superfield has the following expansion:
\begin{equation}
\Phi(x^{\mu}, \theta_{\alpha})=\phi(x^{\mu})+i\bar\theta\psi(x^{\mu})+
\frac{i}{2}\bar\theta \theta F(x^{\mu}),
\end{equation}
where $\phi$ and $F$ are real bosonic (even) 
scalar fields and $\psi_{\alpha}$
is a Majorana spinor field.
The SUSY sine-Gordon equation is:
\begin{equation}\label{ssg}
{\bar D} D\Phi=2i\sin{\Phi},
\end{equation}
where $D_{\alpha}=
\partial_{\theta^{\alpha}}+i(\gamma^{\mu}\theta)_{\alpha}\partial_{\mu}$
and on the components it has the form:
\begin{eqnarray}
(\gamma^{\mu}\partial_{\mu}+\cos{\phi})\psi=0\nonumber\\
\phi_{xx}-\phi_{tt}=\frac{1}{2}(\sin{(2\phi)}-i\bar\psi \psi \sin{\phi}).
\end{eqnarray} 

\section{Super-Hirota operators}

In order to apply the bilinear formalism on these equations one 
has to define a SUSY bilinear operator. We are going to consider the
following general ${\cal N}=1$ SUSY bilinear expression
\begin{equation}\label{susy}
S_{N}(f,g)=\sum_{i=0}^{N}c_{i}(D^{N-i} f)(D^{i} g),
\end{equation}
for any N, where $D$ is 
the covariant derivative and $f$, $g$ are Grassmann valued
functions (odd or even). We shall prove the following

{\it Theorem}: 
The general ${\cal N}=1$ SUSY bilinear expression (\ref{susy})
is super-gauge invariant i.e. 
for $\Theta=kx+\omega t+\theta \hat \zeta+
$...linears ($\zeta$ is a Grassmann 
parameter)
$$S_{N}(e^{\Theta}f,e^{\Theta}g)=e^{2\Theta}S_{N}(f,g),$$
if and only if
$$c_{i}=c_{0}(-1)^{i|f|+\frac{i(i+1)}{2}}\left[  \begin{array}{c}
                                                     N\\i
                                                 \end{array}     \right],$$
where the super-binomial coefficients are defined by:
$$\left[\begin{array}{c}        
              N\\i                        
        \end{array}\right]=                        
\left\{ \begin{array}{ll}
        \left(\begin{array}{c}        
              $[N/2]$   \\   $[i/2]$                        
        \end{array}\right) & \mbox {if $(N,i)\neq(0,1) mod 2$} \\
                         0 & \mbox {otherwise}                      
        \end{array}           
\right. $$
$|f|$ is the Grassmann parity of the function $f$ defined by:
$$|f|=\left\{ \begin{array}{ll}
              1 & \mbox {if $f$ is odd (fermionic)} \\
              0 & \mbox {if $f$ is even (bosonic)}                      
              \end{array}           
\right. $$
and $[k]$ is the integer part of the real number $k$ ($[k]\leq k<[k]+1$)

{\bf Proof:} First we are 
going to consider $N$ even and we shall take it on the
form $N=2P$. In this case we have:
$$
S_{N}(f,g)=\sum_{i=1}^{N}c_{i}(D^{N-i} f)(D^{i} g)=
\sum_{i=0}^{P}c_{2i}(\partial^{P-i} f)(\partial^{i} g)
+\sum_{j=0}^{P-1}c_{2j+1}(\partial^{P-j-1} Df)(\partial^{j} Dg)
$$
Imposing the super-gauge invariance and expanding the covariant 
derivatives we obtain:
$$
\sum_{n\geq 0}\sum_{m\geq 0}\left(
\sum_{i=0}^{P}c_{2i}
\left(  \begin{array}{c}
        i\\n
        \end{array}     \right)
\left(  \begin{array}{c}
        P-i\\m
        \end{array} \right)
k^{P-n-m}\right)(\partial^m f)(\partial ^n g)+$$
$$
+\sum_{n'\geq 0}\sum_{m'\geq 0}\Lambda\left(
\sum_{j=0}^{P-1}c_{2j+1}
\left(  \begin{array}{c}
        j\\n'
        \end{array}     \right)
\left(  \begin{array}{c}
        P-j-1\\m'
        \end{array} \right)
k^{P-n'-m'-1}\right)(\partial^{m'} f)(\partial ^{n'} Dg)+$$
$$
+\sum_{n'\geq 0}\sum_{m'\geq 0}\Lambda(-1)^{|f|+1}\left(
\sum_{j=0}^{P-1}c_{2j+1}
\left(  \begin{array}{c}
        j\\n'
        \end{array}     \right)
\left(  \begin{array}{c}
        P-j-1\\m'
        \end{array} \right)
k^{P-n'-m'-1}\right)(\partial^{m'} Df)(\partial ^{n'} g)+$$
$$
+\sum_{n\geq 0}\sum_{m\geq 0}\left(
\sum_{j=0}^{P-1}c_{2j+1}
\left(  \begin{array}{c}
        j\\n
        \end{array}     \right)
\left(  \begin{array}{c}
        P-j-1\\m
        \end{array} \right)
k^{P-n-m-1}\right)(\partial^m Df)(\partial ^n Dg)=$$
$$
=\sum_{i=0}^{P}c_{2i}(\partial^{P-i} f)(\partial^{i} g)
+\sum_{j=0}^{P-1}c_{2j-1}(\partial^{P-j-1} Df)(\partial^{j} Dg)
$$
where $\Lambda=\hat\zeta+\theta k$. From this, we must have for 
every $m$, $n$ subjected to $0\leq n\leq i\leq P-m$
and $j\leq P-m'$. 
\begin{equation}\label{ddiscret}
\sum_{i=0}^{P}c_{2i}
\left(  \begin{array}{c}
        i\\n
        \end{array}     \right)
\left(  \begin{array}{c}
        P-i\\m
        \end{array} \right)
k^{P-n-m}=c_{2n}\delta_{P-n-m}
\end{equation}
Also due to the fact that the supergauge invariance has to be obeyed
for every $f$ and $g$ we must have $c_{2j+1}=0$ 
The discrete equation (\ref{ddiscret}) was solved in \cite{grammaticos}.
Its general solution is given by:
\begin{eqnarray}\label{par}
c_{2i} &=& c_{0}(-1)^{i}
\left(  \begin{array}{c}
        P\\i
        \end{array}     \right)\nonumber\\
c_{2j+1} &=& 0
\end{eqnarray}

In the case of $N=2P+1$ we proceed in a similar manner and we obtain 
the following system:
\begin{equation}\label{discret}
\sum_{i=0}^{P}c_{2i}
\left(  \begin{array}{c}
        i\\n
        \end{array}     \right)
\left(  \begin{array}{c}
        P-i\\m
        \end{array} \right)
k^{P-n-m}=c_{2n}\delta_{P-n-m}
\end{equation}
\begin{equation}\label{discret}
\sum_{j=0}^{P}c_{2j+1}
\left(  \begin{array}{c}
        j\\n
        \end{array}     \right)
\left(  \begin{array}{c}
        P-j-1\\m
        \end{array} \right)
k^{P-n-m-1}=c_{2n+1}\delta_{P-n-m-1}
\end{equation}
\begin{equation}
(-1)^{|f|}c_{2i}+c_{2i+1}=0
\end{equation}
This system has the following solution:
\begin{eqnarray}\label{impar}
c_{2i} &=& c_{0}(-1)^{i}
\left(  \begin{array}{c}
        P\\i
        \end{array}     \right)\nonumber\\
c_{2i+1} &=& c_{0}(-1)^{i+1+|f|}
\left(  \begin{array}{c}
        P\\i
        \end{array}     \right)
\end{eqnarray}
The relations (\ref{par}), (\ref{impar}) can be written in a compact form
as 
$$c_{i}=c_{0}(-1)^{i|f|+\frac{i(i+1)}{2}}\left[  \begin{array}{c}
                                                     N\\i
                                                 \end{array}     \right].$$
and the theorem is proved.  We mention that the super-bilinear operator
proposed by McArthur and Yung \cite{mcarthur} 
is a particular case of the above super-Hirota
operator.

We shall note the bilinear operator as
$$S_{N}(f,g):={\bf S}_{x}^{N}f\bullet g$$
In the Appendix 1 we list several simple properties of 
this super-Hirota operator.

\section{Bilinear SUSY KdV-type equations}
\vskip 0.5cm
{\it SUSY KdV of Manin-Radul-Mathieu}
\vskip 0.5cm
\noindent In order to use the super-bilinear 
operators defined above we shall consider
the following nonlinear substitution for the superfield:
\begin{equation}
\Phi(t,x,\theta)=2D^{3}\log{\tau(t,x,\theta)}
\end{equation}
where $\tau$ is an even superfield.
Introducing in SUSY KdV (\ref{skdv}) and integrating with respect to $x$ 
we obtain the following 
$$2D\partial_{t}\log{\tau}+3\{(2D^3\log{\tau})(2\partial_{x}^2\log{\tau})\}
+2D^7\log{\tau}=0$$
Using the properties (\ref{A3}), (\ref{A4}), (\ref{A5})
we find
\begin{equation}\label{bbill}
\frac{{\bf S}_{x}{\bf D}_{t}\tau\bullet \tau}{\tau^2}+
3\left(\frac{{\bf S}_{x}^3\tau\bullet \tau}{\tau^2}\frac{{\bf D}_{x}^2
\tau\bullet \tau}{\tau^2}\right)+2\partial_{x}^3(\frac{D\tau}{\tau})=0
\end{equation}
Now we are using the following property of the {\it classical}
Hirota operator \cite{satsuma}
$$\partial_{x}^3(\frac{a}{b})=\frac{{\bf D}_{x}^3 a\bullet b}{b^2}
-3\frac{{\bf D}_{x}a\bullet b}{b^2}\frac{{\bf D}_{x}^2b\bullet b}{b^2}$$
In our case $a=D\tau$ and $b=\tau$. Accordingly
$$2\partial_{x}^3\left(\frac{D\tau}{\tau}\right)=
\frac{{\bf S}_{x}^7\tau\bullet\tau}{\tau^2}-
3\left(\frac{{\bf S}_{x}^3\tau\bullet \tau}{\tau^2}\frac{{\bf D}_{x}^2
\tau\bullet\tau}{\tau^2}\right)$$
Plugging into (\ref{bbill}) we
find the following super-bilinear form
\begin{equation}
({\bf S}_{x}{\bf D}_{t}+{\bf S}_{x}^{7})\tau\bullet \tau=0,
\end{equation}
which is equivalent with  
the form found by McArthur and Yung\cite{mcarthur}
\begin{equation}
{\bf S}_{x}({\bf D}_{t}+{\bf D}_{x}^{3})\tau\bullet \tau=0.
\end{equation}

In order to find the super-soliton 
solutions we are going to use the classical perturbative method
namely the series 
\begin{equation}\label{ser}
\tau=1+\epsilon f^{(1)}+\epsilon^2 f^{(2)}+\epsilon^3 f^{(3)}+...
\end{equation}
where $f^{(i)}$ are even functions.
Equating the power of $ \epsilon$ we find:

\begin{itemize}
\item for $\epsilon$,
\begin{equation}\label{e1}
D(\partial_{t}+\partial_{x}^3) f^{(1)}=0
\end{equation}
\item for $\epsilon^2$,
\begin{equation}\label{e2}
2D(\partial_{t}+\partial_{x}^3) f^{(2)}=-
{\bf S}_{x}({\bf D}_{t}+{\bf D}_{x}^{3})f^{(1)}\bullet f^{(1)} .
\end{equation}
\item for $\epsilon^3$,
\begin{equation}\label{e3}
D(\partial_{t}+\partial_{x}^3) f^{(3)}=-
{\bf S}_{x}({\bf D}_{t}+{\bf D}_{x}^{3})f^{(1)}\bullet f^{(2)} .
\end{equation}
\item for $\epsilon^4$,
\begin{equation}\label{e4}
2D(\partial_{t}+\partial_{x}^3) f^{(4)}=-
2{\bf S}_{x}({\bf D}_{t}+{\bf D}_{x}^{3})f^{(1)}\bullet f^{(3)}
-{\bf S}_{x}({\bf D}_{t}+{\bf D}_{x}^{3})f^{(2)}\bullet f^{(2)}
\end{equation}
and so on.
\end{itemize}
Now if we take $f^{(1)}=e^{kx-k^3 t+\theta\hat\zeta+\eta^{(0)}}$
the equation (\ref{e1}) is satisfied, 
$f^{(2)}=0$, $f^{(3)}=0....$ and the series (\ref{ser}) 
truncates. So, the 1 supersoliton solution is
given by 
\begin{equation}
\tau^{(1)}=1+e^{kx-k^3 t+\theta\hat\zeta+\eta^{(0)}}
\end{equation}
for every $k$ and $\hat\zeta$.

Introducing in the 
super-bilinear equation the 1 soliton solution 
$F=1+\exp{(kx+\omega t+\hat\zeta\theta)}$
one finds the {\it dispersion supermanifold} equation:
$$P(k,\omega,\hat\zeta)\equiv (\hat\zeta+\theta k)(\omega+k^3)=0$$
which imposes $\omega=-k^3$ for every $\hat\zeta$.

In order to find 2 super-soliton solution we take
$f^{(1)}=e^{\eta_1}+e^{\eta_2}$
where $\eta_i=k_i x-k_{i}^{3} t+\theta\hat\zeta_i$
The equation (\ref{e2}) becomes
\begin{equation}\label{e5}
2D(\partial_{t}+\partial_{x}^3) 
f^{(2)}=6k_1 k_2 (k_1-k_2)[(\hat\zeta_1-\hat\zeta_2)+\theta(k_1-k_2)]
e^{\eta_1+\eta_2}
\end{equation}

Taking into account that $\tau$ is a 
Grassmann even function, $f^{(2)}$ must be also even. 
Accordingly, the general solution of (\ref{e5}) 
has the form 
$$f^{(2)}=\left[m_{12}(k_1,k_2,\hat\zeta_1,\hat\zeta_2)+
\theta\hat n_{12}(k_1,k_2,\hat\zeta_1,
\hat\zeta_2)\right]e^{\eta_1+\eta_2},$$ 
$m_{12}$ and $\hat n_{12}$ being Grassmann valued even and odd functions. 
Due to the fact that we have two Grassmann parameters
($\hat\zeta_1$ and  $\hat\zeta_2$)
the {\it most general} 
expressions for $m_{12}$ and $\hat n_{12}$ are given by,
$$m_{12}=(k_1-k_2)^2/(k_1+k_2)^2+\gamma(k_1,k_2)\hat\zeta_1\hat\zeta_2$$
$$\hat n_{12}=a(k_1,k_2)\hat\zeta_1+b(k_1,k_2)\hat\zeta_2$$  
The above forms for $m_{12}$ and $\hat n_{12}$ 
could also be obtained by expanding in 
power series of $\hat\zeta_1$ and $\hat\zeta_2$
and taking into account that in the 
bosonic limit ($\hat\zeta_i\rightarrow 0$),
$m_{12}\rightarrow  (k_1-k_2)^2/(k_1+k_2)^2$ 
(ordinary KdV interaction term)
and $\hat n_{12}\rightarrow 0$

Introducing in the equation (\ref{e5}) we find
$$f^{(2)}=\left[\left(\frac{k_1-k_2}{k_1+k_2}\right)^2
+2\frac{k_1-k_2}{(k_1+k_2)^2}
\hat\zeta_1\hat\zeta_2+
2\theta\frac{(k_1-k_2)
(k_1\hat\zeta_2-k_2\hat\zeta_1)}{(k_1+k_2)^2}\right]e^{\eta_1+\eta_2}$$

Introducing the above forms for $f^{(1)}$ and $f^{(2)}$ in (\ref{e3})
one obtains $f^{(3)}=0$ and then 
$f^{(4)}=0$, $f^{(5)}=0...$ i.e. the series truncates.
So the 2 super-soliton solution is given by;
\begin{equation}
\tau^{(2)}=1+e^{\eta_{1}}+e^{\eta_{2}}+
A_{12}e^{\eta_1+\eta_2}
\end{equation}
where
\begin{equation}
A_{12}=\left(\frac{k_1-k_2}{k_1+k_2}\right)^2+2\frac{k_1-k_2}{(k_1+k_2)^2}
\hat\zeta_1\hat\zeta_2+
2\theta\frac{(k_1-k_2)
(k_1\hat\zeta_2-k_2\hat\zeta_1)}{(k_1+k_2)^2}
\end{equation}
One can easily see that the classical general procedure 
for finding the interaction term given by the equation (\ref{it})
$$A_{12}P(k_1+k_2)+P(k_1-k_2)=0$$ does not work in the SUSY case.

In order to find 3 supersoliton solution we consider
$$f^{(1)}=e^{\eta_1}+e^{\eta_2}+e^{\eta_3}$$
and the equation for $f^{(3)}$ (\ref{e3})becomes:

\begin{equation}
D(\partial_{t}+\partial_{x}^3) f^{(3)}=-
({\bf S}_{x}{\bf D}_{t}+{\bf S}_{x}^{7})
(e^{\eta_1}\bullet A_{23}e^{\eta_2+\eta_3}
+e^{\eta_2}\bullet A_{13}e^{\eta_1+\eta_3}
+e^{\eta_3}\bullet A_{12}e^{\eta_1+\eta_2})
\end{equation}
The solution $f^{(3)}$ of this 
equation (which is very complicated) {\it does not}
cancel the right hand side of 
the equation (\ref{e4}). So, $f^{(4)}$ is not zero and the series
(\ref{ser}) cannot be truncated. 
Accordingly the SUSY KdV equation does not have 3 super-soliton
solution in the standard form for arbitrary $k_i$'s and $\hat\zeta_i$'s.
This seems a quite strange 
paradox, because SUSY KdV is integrable in the sense of Lax.

Anyway, imposing the constraint 
$k_i\hat\zeta_j=k_j\hat\zeta_i$ for every $i$ and $j$
it is easy to prove (see Appendix 2) 
that SUSY KdV possesess the following N-soliton solution
\begin{equation}
\tau^{(N)}=
\sum_{\mu=0,1}\exp{(\sum_{i=1}^{N}
\mu_{i}\eta_{i}+\sum_{i<j}A_{ij}\mu_{i}\mu_{j})},
\end{equation}
where
$$\eta_{i}=k_{i}x-k_{i}^{3} t+\theta\hat\zeta_{i}+\eta_{i}^{(0)}$$
$$\exp{A_{ij}}=\left(\frac{k_{i}-k_{j}}{k_{i}+k_{j}}\right)^{2}$$
$$k_{i}\hat\zeta_{j}=k_{j}\hat\zeta_{i}$$
Solutions with 
constraints on parameters have been found also by Liu and Manas \cite{m},
using SUSY Darboux transformation.
\vskip 1cm

{\it SUSY KdV-B}
\vskip 0.5cm
\noindent This situation is completely different in the SUSY KdV-B case  
\begin{equation}
\Phi_{t}+D^6 \Phi+6D^2\Phi D\Phi=0.
\end{equation}
With the same nonlinear substitution
$$\Phi=2D^3 \log F$$
we obtain the ordinary form
$$({\bf D}_t {\bf D}_x + {\bf D}^{4}_{x})F\bullet F=0$$
which {\it has} N-super-soliton 
solution ({\ref{N}) the 
fermionic contribution being only an additive 
phase i.e. $\eta_{i}=k_{i}x-k_{i}^3 t+\hat\zeta_{i}\theta.$ 
\vskip 0.5cm
{\it SUSY Sawada-Kotera-Ramani}
\vskip 0.5cm
\noindent For ${\cal N}=1$ SUSY 
Sawada-Kotera-Ramani (\ref{sskr}) using the nonlinear
substitution, 
$$\Phi=2D^3\log{\tau(t,x,\theta)}$$
we shall find the following super-bilinear form:
\begin{equation}
({\bf S}_{x}{\bf D}_{t}+{\bf S}_{x}^{11})\tau\bullet \tau=0
\end{equation}

In a similar way we find the 2 super-soliton solution
$$\tau^{(2)}=1+e^{\eta_1}+e^{\eta_2}
+\left[\left(\frac{k_{1}-k_{2}}{k_{1}+k_{2}}\right)^{2}
\frac{k_{1}^2 -k_{1}k_{2}+k_{2}^{2}}{k_{1}^2 +k_{1}k_{2}+k_{2}^{2}}+
2\frac{k_{1}-k_{2}}{(k_{1}+k_{2})^2}
\frac{k_{1}^2 -k_{1}k_{2}+k_{2}^{2}}
{k_{1}^2 +k_{1}k_{2}+k_{2}^{2}}\hat\zeta_1\hat\zeta_2\right]\times$$
$$\times(1+2\theta\frac{k_2\hat\zeta_1
-k_1\hat\zeta_2}{k_1-k_2})e^{\eta_1+\eta_2}$$

Also with the same constraint $k_{i}\hat\zeta_{j}=k_{j}\hat\zeta_{i}$
we find
\begin{equation}
\tau^{(N)}=
\sum_{\mu=0,1}
\exp{(\sum_{i=1}^{N}\mu_{i}\eta_{i}+\sum_{i<j}A_{ij}\mu_{i}\mu_{j})},
\end{equation}
where
$$\eta_{i}=k_{i}x-k_{i}^{5} t+\theta\hat\zeta_{i}+\eta_{i}^{(0)}$$
$$\exp{A_{ij}}=\left(\frac{k_{i}-k_{j}}{k_{i}+k_{j}}\right)^{2}
\frac{k_{i}^2 -k_{i}k_{j}+k_{j}^{2}}{k_{i}^2 +k_{i}k_{j}+k_{j}^{2}}$$
\vskip 0.5cm
{\it SUSY Hirota-Satsuma}
\vskip 0.5cm
\noindent For ${\cal N}=1$ 
SUSY Hirota-Satsuma equation using the nonlinear substitution:
$$\Phi=2D\log{\tau(t,x,\theta)}$$
one obtains the super-bilinear form:
\begin{equation}
({\bf S}_{x}^{5}{\bf D}_{t}-{\bf S}_{x}^3
-{\bf S}_{x}{\bf D}_{t})\tau \bullet \tau=0
\end{equation} 
The 2 supersoliton solution is given by:
$$\tau^{(2)}=1+e^{\eta_1}+e^{\eta_2}
+\left[\left(\frac{k_{1}-k_{2}}{k_{1}+k_{2}}\right)^{2}
M_{12}+2\frac{k_{1}-k_{2}}{(k_{1}
+k_{2})^2}M_{12}\hat\zeta_1\hat\zeta_2\right]
(1+2\theta\frac{k_2\hat\zeta_1
-k_1\hat\zeta_2}{k_1-k_2})e^{\eta_1+\eta_2}$$
where
$$M_{12}=\frac{(k_{1}-k_{2})^2+k_{1}k_{2}
[(k_{1}-k_{2})^{2}-(k_{1}^2-1)(k_{2}^{2}-1)]}
{(k_{1}-k_{2})^2-k_{1}k_{2}[(k_{1}-k_{2})^{2}-(k_{1}^2-1)(k_{2}^{2}-1)]}$$
and $\eta_{i}=k_{i}x-k_{i}t/(k_{i}^2-1)+\hat\zeta_{i}\theta.$
With the constraints 
$k_{i}\hat\zeta_{j}=k_{j}\hat\zeta_{i}$ the equation admits 
also N soliton solution with 
$$A_{ij}=\left(\frac{k_{1}-k_{2}}{k_{1}+k_{2}}\right)^{2}M_{12}.$$ 
\vskip 0.5cm
{\it SUSY Burgers}
\vskip 0.5cm
\noindent An interesting case is the SUSY extension of the 
Burgers equation, namely
$$\Phi_t+\Phi D\Phi_{x}+\Phi_{xx}=0$$
It is welknown 
that the classical 
Burgers equation can be linearized via Cole-Hopf transform.
If we try the nonlinear 
substitution (which is the natural supersymmetrization of the 
Cole-Hopf transform):
$$\Phi=2D\log{\tau(t,x,\theta)}$$
we find:
$${\bf S}_{x}{\bf D}_{t}\tau \bullet \tau 
+2{\bf D}_{x}^2 D\tau \bullet \tau=0.$$
This form is not super-gauge invariant. 
Accordingly we are forced to use 
two functions for substitution, namely
$$\Phi=\frac{\hat G(t,x,\theta)}{F(t,x,\theta)}$$
where $\hat G$ is an 
odd Grassmann function and $F$ is an even one.
Using the relation (\ref{gf}) 
from Appendix 1 we obtain 
the following gauge-invariant super-bilinear form
$$({\bf D}_{t}+{\bf D}_{x}^2)\hat G \bullet F=0$$
$${\bf D}_{x}^2 F\bullet F = {\bf S}_{x}^3 \hat G \bullet F$$
This system admits the following 1 super-shock solution:
$$\hat G=2(\hat\zeta+\theta k)e^{(kx-k^2 t+\hat\zeta \theta)}, 
\quad F=1+e^{(kx-k^2 t+\hat\zeta\theta)}$$
\vskip 1cm

We can ask ourselves if it is possible to obtain super-bilinear forms
for SUSY equations of the nonlinear Klein-Gordon type. In fact 
the SUSY sine-Gordon equation(\ref{ssg})
can be written in the following form:
\begin{equation}
[D_{T}, D_{X}]\Phi(T, X, \theta, \theta_{t})
=2i\sin{\Phi(T, X, \theta, \theta_{t})}
\end{equation}
where we have introduced the light-cone variables
$X:=i(t-x)/2$, $T:=i(t+x)/2$, and 
$\theta:=\theta_{1}$, $\theta_{2}:=-\theta_{t}.$
Covariant derivatives are
$D_{X}:=\partial_{\theta}+\theta\partial_{X}$, 
$D_{T}:=\partial_{\theta_{t}}+\theta_{t}\partial_{T}$
and the square braket means the commutator.
Using the nonlinear substitution ($G$ and $F$ are even functions)
$$\Phi=2i\log{\left(\frac{G}{F}\right)},$$
we find the following quadrilinear expression
$$2i\{F^2
(G[D_{T},D_{X}]G-[D_{T}G, D_{X}G])
-G^2(F[D_{T},D_{X}]F-[D_{T}F, D_{X}F])
\}=F^4-G^4$$
It is easy to see that the bilinear operator
$${\bf S}_{XT}\tau\bullet \tau:
=\tau[D_{T},D_{X}]\tau-[D_{T}\tau, D_{X}\tau]$$
is super-gauge invariant with respect to the super-gauge
$$e^{\Theta}:=e^{(kx+\omega t+\theta\hat\zeta
+\theta_{t}\hat\Omega+liniars)}.$$
Accordingly
we can choose the following super-bilinear form, formally the same with 
standard sine-Gordon equation,
\begin{eqnarray}
{\bf S}_{XT}G\bullet G &=& \frac{1}{2i}(F^2-G^2)\nonumber\\
{\bf S}_{XT}F\bullet F &=& \frac{1}{2i}(G^2-F^2)
\end{eqnarray}
but, it is not clear how to compute the super-kink solutions.

>From these examples it seems that gauge-invariance is a useful concept
for bilinear formalism in the supersymmetric case, 
though there is no deep reason for that. 
As a consequence we was able
to bilinearize two supersymmetric 
equations of KdV type and the SUSY sine-Gordon. The case
of SUSY versions for mKdV, NLS, KP etc. requires further 
investigation because it seems that {\it only
certain supersymmetric extensions are super-bilinearizable}.
Although we do not know if the SUSY extension of 
Sawada-Kotera proposed above
are integrable in the sense of Lax, 
it admits super-bilinear form and only 2 super-soliton
solution for arbitrary choice of solitary waves.
The strange fact is that 
SUSY KDV equation of Mathieu which is known to be Lax integrable 
does not admit $N geq 3$ supersoliton solution in the canonical form.
Probably a singularity
analysis implemented on the super-bilinear form will reveal
the connection between 
Hirota-integrability  and Lax-integrability.
\section{Appendix 1}
In this section we are 
going to list several properties of the super-Hirota
bilinear operator which are useful in deriving bilinear forms.

\begin{equation}\label{A1}
{\bf S}_{x}^{2N}f\bullet g={\bf D}_{x}^{N}f\bullet g
\end{equation}
\begin{equation}\label{A2}
{\bf S}_{x}^{2N+1}e^{\eta_{1}}\bullet e^{\eta_{2}}=
[\hat\zeta_{1}-\hat\zeta_{2}+\theta(k_{1}-k_{2})](k_{1}-k_{2})^{N}
e^{\eta_{1}+\eta_{2}}
\end{equation}
\begin{equation}
{\bf S}_{x}^{2N+1}1\bullet e^{\eta_{1}}
=(-1)^{N+1}(\hat\zeta+\theta k)k^{N}e^{\eta}=
(-1)^{N+1}{\bf S}_{x}^{2N+1}e^{\eta}\bullet 1
\end{equation}
where $\eta_{i}=k_{i}x+\theta\hat\zeta_{i}$ and 
$\hat\zeta_{i}$ are odd Grassmann
numbers.
\begin{equation}\label{A3}
2D\log{\tau}=\frac{D\tau}{\tau}
\end{equation}
\begin{equation}\label{A4}
2D^3\log{\tau}=\frac{{\bf S}_{x}^3 \tau\bullet \tau}{\tau^2}
\end{equation}
\begin{equation}\label{A5}
2D\partial_{t}\log{\tau}=\frac{{\bf S}_{x}
{\bf D}_{t}\tau\bullet\tau}{\tau^2}
\end{equation}
where $\tau$ is an even Grassmann function
Moreover if $G$ and $F$ are Grassmann functions (with F even) then
\begin{equation}\label{gf}
D^{3}(\frac{G}{F})
=\frac{{\bf S}_{x}^{3}G\bullet F}{F^2}-(-1)^{|G|}
\frac{G}{F}
\frac{{\bf S}_{x}^{3}F\bullet F}{F^2}
\end{equation}

\section{Appendix 2}
In this section we are going to sketch the proof of the formula
for N supersoliton solution for bilinear SUSY KdV equation. 
We are rely on the proof of Hirota for ordinary 
N-soliton solution in the case of KdV \cite{hir}, \cite{ablowitz}.
Thus, introducing the expression of the N supersoliton solution
$$\tau^{(N)}=
\sum_{\mu=0,1}\exp{(\sum_{i=1}^{N}
\mu_{i}\eta_{i}+\sum_{i<j}A_{ij}\mu_{i}\mu_{j})},$$
$$\eta_{i}=k_{i}x-k_{i}^{3} t+\theta\hat\zeta_{i}+\eta_{i}^{(0)}$$
$$\exp{A_{ij}}=\left(\frac{k_{i}-k_{j}}{k_{i}+k_{j}}\right)^{2}$$
$$k_{i}\hat\zeta_{j}=k_{j}\hat\zeta_{i}$$
into the super bilinear form 
$$({\bf S}_{x}{\bf D}_{t}+{\bf S}_{x}^{7})\tau\bullet \tau=0,$$
and taking into account the properties
(\ref{A1}) and (\ref{A2}) we find
$$\sum_{\mu=0,1}\sum_{\nu=0,1}\{
(\sum_{i=1}^N(\mu_{i}-\nu_{i})\Lambda_{i})
((-)\sum_{i=1}^N(\mu_{i}-\nu_{i})k_{i}^3)+
(\sum_{i=1}^N(\mu_{i}-\nu_{i})\Lambda_{i})
(\sum_{i=1}^N(\mu_{i}-\nu_{i})k_{i})^3\}\times$$
$$\times\exp{(\sum_{i=1}^{N}(\mu_{i}+\nu_{i})\eta_{i}
+\sum_{i<j}(\mu_{i}\mu_{j}+\nu_{i}\nu_{j})A_{ij})}=0$$
where 
$\Lambda=\hat\zeta_{i}+\theta k_{i}$.
Since $\mu_{i}, \nu_{i}=0,1$ it is 
clear that we only have exponential terms
of the form
$$\exp{(\sum_{i=1}^n \eta_{i}+\sum_{i=n+1}^m 2\eta_{i})}, 0\leq n \leq N, 
n \leq m \leq N $$
Next we show that 
the coefficient of this general exponential term is zero,
the coefficient being given by, \cite{hir}\cite{ablowitz}:
\begin{equation}\label{delta}
\Delta=\sum_{\sigma=0,1}\{
(-\sum_{i=1}^N\sigma_{i}\Lambda_{i})
(\sum_{i=1}^N\sigma_{i}k_{i}^3)+
(\sum_{i=1}^N\sigma_{i}\Lambda_{i})
(\sum_{i=1}^N\sigma_{i}k_{i})^3\}
\prod_{i<j}^{n}(\sigma_{i}k_{i}-\sigma_{j}k_{j})^2
\end{equation}
where $\sigma_{i}=\mu_{i}-\nu_{i}$.
We do not go into details because  
the procedure of deriving the 
above coefficient goes absolutely in the same way
as in the case of ordinary KdV equation. This is due to the fact that
the interaction term $A_{ij}$ is the same for KdV and SUSY KdV.

Because $\Lambda=\hat\zeta_{i}+\theta k_{i}$ the coefficient
$\Delta$ becomes on the components
\begin{equation}
\Delta\equiv\Delta_{0}+\Delta_{1}=\sum_{\sigma=0,1}\{
(-\sum_{i=1}^N\sigma_{i}\hat\zeta_{i})
(\sum_{i=1}^N\sigma_{i}k_{i}^3)+
(\sum_{i=1}^N\sigma_{i}\hat\zeta_{i})
(\sum_{i=1}^N\sigma_{i}k_{i})^3\}
\prod_{i<j}^{n}(\sigma_{i}k_{i}-\sigma_{j}k_{j})^2+
\end{equation}
$$+\theta\sum_{\sigma=0,1}\{
(-\sum_{i=1}^N\sigma_{i}k_{i})
(\sum_{i=1}^N\sigma_{i}k_{i}^3)+
(\sum_{i=1}^N\sigma_{i}k_{i})
(\sum_{i=1}^N\sigma_{i}k_{i})^3\}
\prod_{i<j}^{n}(\sigma_{i}k_{i}-\sigma_{j}k_{j})^2$$
Hirota proved that the second term is zero \cite{hir}, \cite{ablowitz}.
The first term is also zero because, using the property
$\hat\zeta_{i}k_{j}=\hat\zeta_{j}k_{i}$, it can be written
as $$\Delta_{0}=\frac{\hat\zeta_{m}}{k_{m}}\Delta_{1}=0$$


\begin{thebibliography}{99}
\bibitem{di vecchia}P. di Vecchia, 
S. Ferrara, Nucl. Phys. {\bf B 130}, 93, (1977)
\bibitem{chaichain}M. Chaichain, 
P. Kulish, Phys. Lett. {\bf B 78}, 413, (1978)
\bibitem{manin}Yu. Manin, A. Radul, 
Comm. Math. Phys. {\bf 98}, 65, (1985)
\bibitem{mathieu}P. Mathieu, J. Math. Phys. {\bf 29}, 2499, (1988)
\bibitem{beck} K. Becker, M. Becker, 
Mod. Phys. Lett {\bf A8}, 1205, (1993)
\bibitem{yung}C. M. Yung, Phys. Lett. {\bf B309}, 75, (1993)
\bibitem{oevel}W. Oevel, Z. Popowicz, 
Comm. Math. Phys. {\bf 139}, 441, (1991)
\bibitem{ueno}K. Ueno, H. Yamada, 
Adv. Stud. in Pure Math. {\bf 16}, 373, (1988)
\bibitem{medina}L. A. Ibort, L. Martinez Alonso, 
E. Medina, J. Math. Phys. {\bf 37}, 6157, (1996)
\bibitem{m}Q. P. Liu, M. Manas, solv-int/9806005
\bibitem{alvarez}L. Alvarez-Gaume, H. Itoyama, 
J. L. Manes, A. Zadra, Int. Journ. Mod. Phys. {\bf A 7}, 5337, (1992)
\bibitem{kac}V. Kac, J van de Leur, 
Ann. de L' Inst. Fourier, {\bf 37}, 99, (1987)
\bibitem{kacmed}V. Kac, E. Medina, Lett. Math. Phys. {\bf 37}, 435(1996)
\bibitem{leclair}A. LeClair, Nucl. Phys. {\bf B314}, 435, (1989)
\bibitem{mcarthur}I. N. McArthur,
 C. M. Yung, Mod. Phys. Lett. {\bf A 8}, 1739, (1993)
\bibitem{fane}A. S. Carstea, 
solv-int/9812022, {\it to appear in}  Nonlinearity
\bibitem{grammaticos}B. Grammaticos, A. Ramani, 
J. Hietarinta, Phys. Lett. {\bf A 190}, 65, (1994)
\bibitem{sw}K. Sawada, T. Kotera, Progr. Theor. Phys. {\bf 51}, 1355, (1974)
\bibitem{satsuma}R. Hirota, J. Satsuma, Progr. Theor. Phys. Suppl.
, {\bf 59}, (1976)
\bibitem{hirota1976}R. Hirota, J. Satsuma, 
J. Phys. Soc. Jap. {\bf 40}, 611, (1976)
\bibitem{hir}R. Hirota, Phys. Rev. 
Lett. {\bf 27}, 1192, (1971)
\bibitem{hieta}J. Hietarinta, "Hirota's bilinear
 method and partial integrability" in 
{\it Partially integrable evolution equation in physics} ed.
 R. Conte, N. Boccara, Kluwer (1990) 
\bibitem{ablowitz}M. J. Ablowitz, H. Segur, 
{\it Solitons and the Inverse Scattering Transform}
SIAM, Philadelphia, (1981)
\bibitem{inami}Inami, Kanno, Comm. Math. Phys. {\bf 136}, 519, (1991)
\bibitem{tsy}P. Kulish, S. Tsyplyaev, 
Theor. Math. Phys. {\bf 46}, 172, (1981)
\end{thebibliography}
\end{document}